\documentclass[10pt,journal,compsoc]{IEEEtran}
\ifCLASSOPTIONcompsoc
  \usepackage[nocompress]{cite}
\else
  \usepackage{cite}
\fi

\usepackage{graphicx}
\usepackage{balance}
\usepackage{multicol}
\usepackage{acronym}
\usepackage{soul}
\usepackage{xcolor}
\usepackage[T1]{fontenc}

\acrodef{D-D}[D-D]{Decisions~\&~Disruptions}
\acrodef{GDPR}{General Data Protection Regulation}
\acrodef{SME}{Small and Medium-sized Enterprise}
\acrodef{CISO}{Chief Information Security Officer}
\acrodef{SCADA}{Supervisory Control and Data Acquisition}
\acrodef{NPO}{National Public Organization}
\acrodef{BCSG}{Bristol Cyber Security Group}

\newcommand{\header}[1]{\multicolumn{1}{c}{#1}}
\usepackage{amsmath}

\newcommand{\overunder}[2]{\ensuremath{\genfrac{}{}{0pt}{}{\text{#1}}{\text{#2}}}}
\newcommand{\mean}[1]{\ensuremath{\genfrac{}{}{0pt}{}{\text{average}}{\text{#1}}}}

\def\tablestyle{\footnotesize\centering}

\def\thegap[0]{\baselineskip}
\usepackage{xparse}
\newcommand{\quotesource}[1]{\hspace*{\fill}{\scriptsize{---#1}}}
\NewDocumentEnvironment{ddquote}{o}{
  \begin{center}
  \begin{minipage}{\dimexpr \linewidth - 2em}\noindent\itshape\small\sffamily
  }{
    \IfNoValueF{#1}{\upshape\quotesource{#1}}
    \end{minipage}
    \end{center}
  }

\usepackage{longtable}
\usepackage[hidelinks]{hyperref}
\usepackage{microtype}
\PassOptionsToPackage{final}{microtype}

\usepackage{booktabs}

\newsavebox{\fminipagebox}
\NewDocumentEnvironment{fminipage}{m O{\fboxsep}}
 {\par\kern#2\noindent\begin{lrbox}{\fminipagebox}
  \begin{minipage}{#1}\ignorespaces}
 {\end{minipage}\end{lrbox}%
  \makebox[#1]{%
    \kern\dimexpr-\fboxsep-\fboxrule\relax
    \fbox{\usebox{\fminipagebox}}%
    \kern\dimexpr-\fboxsep-\fboxrule\relax
  }\par\kern#2
 }

\NewDocumentEnvironment{takeaway}{}{\begin{fminipage}{\linewidth}\sffamily\small\textbf{Take-away:}~}{\end{fminipage}}

\usepackage{subcaption}

\begin{document}

\title{The best laid plans or lack thereof: Security decision-making of different stakeholder groups}

\author{Benjamin~Shreeve, Joseph Hallett, Matthew Edwards, Kopo M. Ramokapane, Richard Atkins and Awais Rashid~\IEEEmembership{Member,~IEEE}%
\IEEEcompsocitemizethanks{\IEEEcompsocthanksitem B, Shreeve, J. Hallett, M. Edwards, K. Ramokapane and A. Rashid are with the \ac{BCSG}, University of Bristol, UK.\protect\\
E-mail: \{ben.shreeve, joseph.hallett, matthew.john.edwards, marvin.ramokapane, awais.rashid\}@bristol.ac.uk
\IEEEcompsocthanksitem R. Atkins is with the City of London Police, UK.}
\thanks{Manuscript received December 15, 2019; revised XXX.}}

\IEEEtitleabstractindextext{
\begin{abstract}
Cyber security requirements are influenced by the priorities and decisions of a range of stakeholders.
Board members and \acp{CISO} determine strategic priorities.
Managers have responsibility for resource allocation and project management.
Legal professionals concern themselves with regulatory compliance. Little is understood about how the security decision-making approaches of these different stakeholders contrast, and if particular groups of stakeholders have a better appreciation of security requirements during decision-making. Are risk analysts better decision makers than \acsp{CISO}? Do security experts exhibit more effective strategies than board members? This paper explores the effect that different experience and diversity of expertise has on the quality of a team's cyber security decision-making and whether teams with members from more varied backgrounds perform better than those with more focused, homogeneous skill sets.
Using data from 208 sessions and 948 players of a tabletop game run \emph{in the wild} by a major national organization over 16 months, we explore how choices are affected by player background (e.g.,~cyber security experts versus risk analysts, board-level decision makers versus technical experts) and different team make-ups (homogeneous teams of security experts versus various mixes). We find that no group of experts makes significantly better game decisions than anyone else, and that their biases lead them to not fully comprehend what they are defending or how the defenses work.

\end{abstract}}

\maketitle


\section{Introduction}
\label{sec:introduction}

\IEEEPARstart{T}{he} cyber security of any system is heavily impacted by the decisions of various stakeholders (e.g.~\acp{CISO}, other board members, managers, engineers, HR and legal teams).
Each stakeholder has differing security requirements, and prioritize their needs differently. Numerous technical solutions can mitigate some risks~\cite{abbasi-2017-ecfi,carcano-2011-multidimensional,chen-2018-learning,eldefrawy-2012-smart}; and various risk and asset-management techniques can guide their deployment~\cite{iso27001,nist-800-53}---but it is the stakeholders who decide what security requirements to prioritize. Economics plays a role in this process, as not all requirements can be met, given the need to balance investment in security against core products and services.
Such decisions are not just within the purview of security specialists. Decisions of a range of other stakeholders are key in determining security requirements and their prioritization. Board members, for instance, set strategic priorities for the organization. Managers hold the resources that need to be allocated to service a range of requirements, not just security. IT personnel are responsible for day-to-day operations of the organization's infrastructure, developing, procuring and deploying a range of in-house and third party services (that may impact security). Legal experts decide on prioritization with regards to regulatory compliance for a range of aspects, including security \& privacy. Physical security personnel are charged with protecting the physical assets, an important consideration in protecting direct access to an organization's connected infrastructure. In this paper, we investigate how these various stakeholder groups complete a cyber security investment game, what do they prioritize and do any teams perform better than others in terms of game outcome?

To explore decision-making in these scenarios Frey~et~al{.} created the \acf{D-D} game~\cite{Frey2019TheGoodBadandUgly}. \ac{D-D} challenges teams to defend a hydro-electric power generation facility from attack, by deciding which of multiple defenses they should invest in, with a finite budget over four rounds.
The game was released under a \emph{CC~By-NC~4.0} license\footnote{\url{www.decisions-disruptions.org}}. It has been updated and adopted as a training and awareness tool for public and private sector organizations by a large \ac{NPO}.

Using their two most recent versions of the game (which add financial punishments and reflect the introduction of the \ac{GDPR}), the \ac{NPO} has played 208 games nationwide with organizations including branches of the police, government departments, banks, \acp{SME} and large multi-national businesses. These games (lasting 60--90 minutes) include teams composed of mixed player backgrounds as well as teams with only one kind of experts. In this paper, we present an analysis of the data gathered using these two different versions of \ac{D-D} run \emph{in the wild} by the \ac{NPO}.
Specifically, we analyze the round-by-round choices and game outcomes from all 208 games, as well as transcriptions of the in-game discussion of 8 teams that we were allowed to record by the \ac{NPO}\footnote{13 additional recordings were untranscribable due to noise.}. The goal of our analysis is to identify patterns in teams' decision-making processes in order to answer three research questions (RQs):

\begin{description}

\item[\textbf{RQ1.}] What effect does a player's background have on their game performance?

\item[\textbf{RQ2.}] What is the effect of diversity of expertise on a team's performance?

\item[\textbf{RQ3.}] What patterns can be seen in the decision-making processes of players and teams?

\end{description}

We begin with a quantitative analysis of the decisions across all games, followed by a qualitative analysis of the transcripts of 8 teams for which we have recordings. 
We find that no stakeholder group is better at making cyber security decisions compared to others (i.e., makes cyber security investment choices that result in the organization suffering significantly fewer cyber security incidents).
Specialist cyber security experts play no better than risk experts, managers, or board-level executives.
Public and private sectors perform no better than other groups. Mixed teams of experts do no better than homogeneous ones.

We then present a thematic analysis of the 8 transcripts available to us. These transcripts happen to be of 8 homogeneous teams of security experts. The analysis highlights several decision-making patterns of security experts. Contrasting this qualitative analysis with the decisions made by the full set of 208 teams indicates potential underpinning causes as to \emph{why} different stakeholders make the security decisions they do and \emph{why} security experts do no better than other stakeholder groups:

\textbf{\textit{Lack of systematic threat modeling or risk analysis.}} Players make decisions not necessarily based on any particular threat or risk framework (only 1 team employed any form of systematic threat modeling), but instead base decisions on superstition, gut instinct, and a poor understanding of how the technology actually works. Furthermore, players disregarded threat intelligence and risk information when it was provided through the game, and tended to conflate key concepts such as threats and vulnerabilities.

\textbf{\textit{Lure of familiar security solutions and technology panaceas.}} Players use well-known security tools, such as \emph{Firewalls} and \emph{Antivirus}, as an alternative to experience, deploying them without truly understanding what it is they defend and why they are used.
They also preferred buying new kit to defend the infrastructure instead of upgrading or patching the technology they already had in place.

\textbf{\textit{Knee-jerk vs. laissez-faire responses to attacks.}} Players typically did not rationalize the attacks they suffered. Consequently, their responses were either \emph{knee-jerk}, and hence costly whilst leaving other security requirements de-prioritized, or \emph{laissez-faire}, disregarding this new contextual information and continuing with their prior decisions.

Our study is the first to explore how security experts compare with stakeholders from a range of different backgrounds in a large dataset collected in the wild by an \ac{NPO}. Our findings confirm some of Frey~et~al{.}'s~\cite{Frey2019TheGoodBadandUgly}\textemdash security experts do not perform better than other stakeholders, and players exhibited the \emph{tunnel vision syndrome}, disregarding in-game information and not questioning prior decisions. However,
any comparisons with Frey~et~al{.}'s work must be made with the understanding that, while the games are similar, they are not the same. The versions reported here have been specifically designed by the large \ac{NPO} to reflect the security landscape as they see it. Additionally, our dataset contains two different variants of the game---75 games played with a version that introduced penalties, and 133 games played with a later \ac{GDPR} version. We have merged these two datasets as the differences between how teams played these games are not significant (Section~\ref{sec:differences-between-game-versions}). Furthermore, Frey~et~al{.}'s findings are based on a qualitative analysis of 12 games of three stakeholder types (security experts, managers, computer scientists) playing in homogeneous teams. In contrast, ours is a large dataset of 208 games with 12 stakeholder types (Table~\ref{tab:skills}) playing the game in both homogeneous and heterogeneous teams. The fact that both studies identify some similar decision-making patterns highlights the need to help decision-makers understand their own biases when it comes to making security decisions and their relative prioritization---encouraging them to ask \emph{why} they are making particular decisions and ensure that this is not just because the choices are familiar or offer one-stop solution to all security problems.

\section{Background and dataset}

\subsection{Decisions \& Disruptions (D-D)}

\ac{D-D} is a tabletop role-playing game about security decisions in cyber-physical systems
~\cite{Frey2019TheGoodBadandUgly}. The game was designed based on extensive experience of building and developing a cyber-physical systems security testbed~\mbox{\cite{green2017pains}}, analysis of major existing cyber security incidents (at the time) impacting critical infrastructures~\mbox{\cite{frey2016role}} and insights from fieldwork in such environments~\mbox{\cite{zanutto2017shadow}} (in addition to drawing upon various collaborative discussions with practitioners in multiple research projects).
The game challenges teams of 5--8 players to help a small utility company improve their current state of security. Teams are informed that the organization operates from two locations: 1) the main production (or plant) site, where a river turns turbines and generates electricity; and 2) an office site which is located elsewhere. The two sites are connected via the internet through a router (but no Firewall), and each site has its own local area network with some computers and production databases. The production site also has a \ac{SCADA} controller connected to the network which runs the turbines. On the office site, a server runs the company's website and email system.

The game is run by a \emph{game master}, who informs teams that they have several different technologies and approaches in which they can invest to secure the company (Table~\ref{tab:investment-costs}). The \emph{game master} follows a strict script, only using prescribed descriptions in order to ensure that all teams have the same decision-making experience. The game takes place over 4 rounds with teams given a finite budget of \$100K in each round (teams are informed that prices are in game \$ and not meant to reflect real-world costs of similar defenses). Teams are able to roll-over remaining funds between rounds. At the end of each round players are told about the attacks they have suffered, and the ones they managed to defend as a result of their choices. All attacks are predetermined, triggered by specific investment choices to ensure teams make their choices under consistent circumstances.
The original version of \mbox{\ac{D-D}} was tested with teams of cyber security practitioners in order to arrive at a scenario which was simple enough to be played by a wide range of stakeholders, but which still reflected some of the key cyber security investment challenges and decisions faced by stakeholders in organizations~\mbox{\cite{Frey2019TheGoodBadandUgly}}.

\begin{table}
  \tablestyle
  \caption{Cost of each investment in the game.  Items with a $\star$ are only available \emph{after} the \emph{Asset Audit} has been played.}
  \newcommand{\money}[1]{\$#1,000}
  \newcommand{\aastar}[1]{#1$^\star$}
  \newcommand{\ditto}[0]{---\textquotedbl---}
  \begin{tabular}{p{0.8\linewidth}c}
    \toprule
    \header{Investment}             & \header{Cost} \\
    \midrule
    \noindent\parbox[c]{\linewidth}{\raggedright \aastar{Encryption (PCs)} / \aastar{Encryption (Databases)} / Threat~Assessment}        & \money{20}    \\\addlinespace
    \noindent\parbox[c]{\linewidth}{\raggedright Antivirus / Asset Audit / \aastar{Controller Upgrade} / Firewall (Office) / Firewall (Plant) / \aastar{PC Upgrade} / Security Training / \aastar{Server Upgrade}}                     & \money{30}    \\\addlinespace
    \noindent\parbox[c]{\linewidth}{\raggedright CCTV (Office) / CCTV (Plant) / Network Monitoring (Office) / Network Monitoring (Plant)}                 & \money{50}    \\
    \bottomrule
  \end{tabular}
  \label{tab:investment-costs}
\end{table}

\subsection{Dataset}

The dataset for this paper comes from two modified and updated versions of the game that were developed by the \ac{NPO}, and which aligns the game's priorities with those of the Cyber Essentials Scheme\mbox{~\cite{ncsc-cyber-essentials}}---a UK Government program demonstrated to be effective at raising the basic level of cyber security within \mbox{\acp{SME}}~\mbox{\cite{such2019basic}}, whose controls are included within NISTIR~7621~\mbox{\cite{nistir-7621}} and which is similar to other European schemes~\mbox{\cite{enisa2016review}}.

The modifications change the game master's script and add explicit financial penalties when teams suffer attacks at the end of each round. The investment options available to participants have not changed, nor the layout of the infrastructure and general premise of the game. However, in the new versions, teams receive different responses to their actions, some of which include financial penalties. For example, in the original version of the game, teams that did not invest in the \emph{Antivirus} option in the first round would receive no penalty, while in the versions analyzed in our study participants would be informed that:

\begin{ddquote}
    ``Ransomware has somehow made its way onto your networks. All computers at your plant and office have been locked. You decide to pay the ransom, a total of \$5,000. Luckily, this unlocks all the computers and you get all your data back.''
\end{ddquote}

\noindent The scenarios have been changed to reflect the attackers and attack vectors which the \ac{NPO} considers to be the most pertinent for the players to learn. The teams reported in this paper played the game using either a \emph{pre-\ac{GDPR}} version of the game or \emph{post-\ac{GDPR}}. The \emph{post-\ac{GDPR}} version of the game represents a minor adjustment to some of the consequences teams encounter, reflecting the change in legislation. Such changes relate to attacks on the database and computers from where sensitive information could be stolen. For example, in the \emph{post-\ac{GDPR}} iteration of the game, if teams have not deployed a \emph{Firewall} on their plant site by the end of the second round, they receive the following warning:

\begin{ddquote}
  ``The white hat Information Security student [who attacked them in round 1 for lack of a firewall] is bored one day and decides to see if you now have basic cyber security at your plant. They're astounded to see that you still don't have a Firewall. They report you directly to the ICO. The ICO inform you that they consider your previous breach an aggravating factor. They decide to issue you a fine of \$150K. They warn you that any further breaches, which are a risk to the rights and freedoms of data subjects will not be tolerated.''
\end{ddquote}

\noindent Seven of the attacks that teams suffer have been adjusted with the introduction of the \emph{post-\ac{GDPR}} version of the game (see Table~\ref{tab:D-D_version_differences}). In each instance the description of the attack suffered for failing to invest in a particular defense (by a particular point in the game) has been updated to reflect \ac{GDPR} legislation. The core scenario used in each description remains the same as the \emph{pre-\ac{GDPR}} version of the game. Small adjustments to the fines have been made to improve game-play, while two fines have been significantly increased to reflect the severity of potential penalties under \ac{GDPR}.

\begin{table}
  \tablestyle
  \caption{Summary of differences in attacks suffered between \emph{pre-\ac{GDPR}} and \emph{post-\ac{GDPR}} iterations of \ac{D-D}}
  \begin{tabular}{r r p{0.45\linewidth}}
    \toprule
    \header{Investment} & \header{Round} & \header{Difference} \\
    \midrule
    Asset audit & 2 & Description \\
    Database Encryption & 3 & Description \\
    Database Encryption & 4 & Description \\
    Firewall (Plant) & 1 & Description, \$12K $\rightarrow$ \$20K \\
    Firewall (Plant) & 2 & Description, \$50K $\rightarrow$ \$150K \\
    PC Upgrade & 4 & Description, \$40K $\rightarrow$ \$50K \\
    Server Upgrade & 4 & Description, \$1M $\rightarrow$ \$5M \\
    \bottomrule
  \end{tabular}
  \label{tab:D-D_version_differences}
\end{table}

The \mbox{\ac{NPO}} also stipulated that the GM should direct players to read the descriptions printed on the front of the \mbox{\emph{Asset}} \mbox{\emph{Audit}} and \mbox{\emph{Threat}} \mbox{\emph{Assessment}} cards. They recognized that the terms \emph{Asset Audit}, \emph{Penetration Testing} and \emph{Threat Assessment} (amongst others) are often used interchangeably to mean similar things---teams were therefore asked to read the two cards carefully to get a better idea of what information they would provide. The \emph{Asset Audit} reads: ``The entire infrastructure is thoroughly assessed for vulnerabilities''---such a process is often completed as part of ISO 27035~\mbox{\cite{iso27035}}, or NIST 800-61~\mbox{\cite{nist-800-61}}. The \emph{Threat Assessment} reads: ``Reveals existing threats to the company, the attack vectors they use, and the possible effects of their attacks''.

Our dataset was collected by the \ac{NPO}, which ran games at companies, workshops, and conferences. The dataset includes 208 games of \emph{pre}~and~\emph{post-\ac{GDPR}} \ac{D-D} played with 948 participants and comprises basic information about the teams (job titles, organization sectors and number of participants) as well as a round-by-round breakdown of the investments they selected.  The \ac{NPO} also allowed us access to audio recordings for 21 of the games (Institutional Review Board approval was provided). The discussions from 8 games were  transcribed---unfortunately noise and cross-talk from the rooms where the games were run meant that the remaining 13 recordings were incomprehensible.

\subsection{Differences between game versions}
\label{sec:differences-between-game-versions}

The question arises as to whether the behaviour of teams differs across the two versions of the game in our data. Table~\ref{tab:round_pos} presents information on the decisions made by teams and the average round position of each decision.
To account for whether the difference in proportions of teams making investment decisions could have arisen by chance, we use a two-sided test of the equality of proportions. We fail to reject the null that the proportion is the same between the two versions ($p=0.58$).
A series of 15 Bonferroni-corrected (accounting for the multiple tests) two-sample t-tests (with $\alpha=0.05$) failed to reject the null that average round positions were equivalent for any decision.

Comparisons between count and round position, along with the small average position differences ($\delta$ in Table~\ref{tab:round_pos}) demonstrates that the behaviour of teams is not different between versions. As the versions make no alterations to causes of \emph{bust} outcomes, these also do not significantly differ (31\% and 28\%, $p=0.72$). However, the differences in losses awarded for attacks between the two versions mean that overall scores of non-bust teams do differ significantly (using Cohen's $d$ to measure effect size: $d=0.35, p=0.04$). As such, while analyses relating to decisions and bust state can be grouped, results relating to non-bust team scores will be confirmed in the two version sub-populations.

\begin{table}
\tablestyle
\caption{Count of teams making investments and the average round position of those investments, for both game versions. Differences in counts and positioning were not significant.}
\begin{tabular}{l rr rr r}
  \toprule
  \multicolumn{1}{c}{}                              & \multicolumn{2}{c}{V1} & \multicolumn{2}{c}{GDPR} & \\
  \header{Decision} & \header{count} & \header{pos.} & \header{count} & \header{pos.} & \header{$\delta$} \\
  \midrule
  Firewall (Plant)		&  75   & 1.25 & 133   & 1.29 & 0.04 \\
  Threat Assessment 		&  72   & 1.54 & 128   & 1.46 & -0.08 \\
  Firewall (Office) 		&  75   & 1.45 & 133   & 1.53 & 0.07 \\
  Antivirus 			&  71   & 1.68 & 125   & 1.72 & 0.04 \\
  Asset Audit 			&  74   & 2.46 & 134   & 2.07 & -0.38 \\
  Security Training 		&  74   & 2.00 & 133   & 2.20 & 0.20 \\
  Encryption (Databases)	&  69   & 2.96 & 126   & 2.60 & -0.35 \\
  Controller Upgrade 		&  69   & 3.01 & 129   & 2.84 & -0.17 \\
  CCTV (Plant) 			&  68   & 3.13 & 113   & 2.97 & -0.16 \\
  Server Upgrade 		&  65   & 3.03 & 124   & 3.12 & 0.09 \\
  Net. Monitoring (Office) 	&  42   & 3.17 &  66   & 3.24 & 0.08 \\
  PC Upgrade 			&  49   & 3.31 &  99   & 3.29 & -0.01 \\
  Encryption (PCs) 		&  52   & 3.40 & 100   & 3.40 & -0.00 \\
  Net. Monitoring (Plant) 	&  38   & 3.00 &  77   & 3.45 & 0.45 \\
  CCTV (Office) 		&  27   & 3.56 &  36   & 3.50 & -0.06 \\
\bottomrule
\end{tabular}
\label{tab:round_pos}
\end{table}

\section{Method}
\label{sec:idea}

\noindent\textbf{RQ1. What effect does a player's background have on their game performance?}

{\setlength{\parskip}{0pt}We hypothesize that participants with specific backgrounds in risk management or cyber security would perform better in a game designed to test these skills. To answer this, we categorize the backgrounds of the various players of each game 
based on their job title and organization.

In order to create these categories, two authors worked through the list of teams with a representative from the \ac{NPO}. They discussed each participant's job title and organization---the \ac{NPO} provided additional clarification on the organizations and participants involved in each session (because the authors were not present during these sessions). In total, 12 distinct categories of player background were identified (Table~\ref{tab:skills}) and 8 categories that described the different organizations where the players worked (Table~\ref{tab:backgrounds}).
We randomly selected 10\% of the mappings, which were remapped by a third author. \emph{Cohen's kappa}~\cite{cohen1960coefficient} was 0.842, which suggests high inter-rater reliability.

We measure game performance in terms of the two outcome measures provided by the game mechanics:
whether or not a team suffered a critical failure and went \emph{bust} during their session; and
the average score (reported as losses suffered) among non-bust teams.
We explore hypotheses about group differences, and use multiple regression to estimate the coefficients for each skill and background in terms of both outcomes.}\\

\begin{table}
  \tablestyle
  \caption{Categorization of player background and a count of their frequency within our dataset, in terms of unique players and teams containing such members (105 teams (50\%) were mixed groups).}
  \begin{tabular}{p{0.2\linewidth} rr p{0.42\linewidth}}
    \toprule
    \header{Skill}&\header{Players}&\header{Teams}&\header{Description} \\
    \midrule
    Board                     & 65	& 44    & Company executives and \newline directors (e.g., CEO, CISO) \\
    Cyber security specialist & 260	& 80    & Cyber security specialists (e.g., consultants, pen testers) \\
    Financial                 & 23	& 14    & Person working in the finance sector (e.g., accountant) \\
    Government                & 21	& 11    & Person who works within government (e.g., politician, civil servant) \\
    Legal                     & 20	& 15    & Person working primarily within the legal system (e.g., lawyer) \\
    Management                & 67	& 38    & A generic manager (e.g., HR manager) \\
    Other                     & 87	& 36    & The rest  (e.g., interns) \\
    Physical \newline Security         & 5	& 5     & Physical security specialists or practitioners (e.g., guards) \\
    Police                    & 129 	& 40    & Police officers or forensic investigators \\
    Risk specialist           & 122	& 62    &  Risk-related workers without cyber security specific experience (e.g., auditors) \\
    Student                   & 103	& 21    & Students and apprentices (with cyber security or forensics focus) \\
    Technical                 & 46 	& 35    & Person who manages an organization's technical infrastructure (e.g., IT personnel, systems administrator) \\
    \midrule
    \emph{Total}              & 948	& 208 	& \\
    \bottomrule
  \end{tabular}
  \label{tab:skills}
\end{table}

\begin{table}
  \tablestyle
  \caption{Categorization of the organizations and a count of their frequency} 
  \begin{tabular}{l r l}
    \toprule
    \header{Sector} & \header{Count} & \header{Description} \\
    \midrule
    Public Sector    & 29 & Public sector orgs. and charities         \\
    Private Sector   & 80 & Private sector businesses                         \\
    Academia         & 19 & Universities                                      \\
    \acs{SME}        & 3  & Smaller orgs.                            \\
    Business Support & 10 & Orgs. providing business support \\
    Mixed            & 23 & Mixed sector teams  \\
    Police           & 39 & Law enforcement                                   \\
    Unknown          & 5  & Unknown organization                              \\
    \bottomrule
  \end{tabular}
  \label{tab:backgrounds}
\end{table}

\noindent\textbf{RQ2. What is the effect of diversity of expertise on a team's performance?}

{\setlength{\parskip}{0pt}If individuals have different perspectives about risk, then we might hypothesize that teams with a more diverse make up might make better risk decisions as a whole, as they ought to be able to consider a greater number of possible attacks.  Approaches to dealing with risk, such as threat modeling, require careful consideration of different possible attacks~\cite{myagmar2005threat,kotonya1998requirements}---a more diverse team should be able to consider a wider range of them.  In \ac{D-D} it is not possible to mitigate all possible attacks by deploying all possible mitigations in a single round, so effective decision making is required to ensure the \emph{appropriate} mitigations are deployed. Again, one may reasonably expect that teams comprised of individuals with a range of different expertise do better than teams with, for instance, only board-level decision makers, technical security experts, or other homogeneous compositions.

For each team we calculate the Shannon index (a measure of diversity)~\cite{spellerberg2003tribute} for the team composition and test the correlation of this measure with the final score and bust state game outcomes. We also explore the difference in outcomes between homogeneous teams and heterogeneous teams of any richness, as these groups bisect our data.}

\begin{table}[t]
  \tablestyle
\caption{Breakdown of game outcomes by player background of teams (identified as any member having such a background), with test statistics and uncorrected p-values. No subgroups were significantly different from the overall means at $\mathbf{\alpha=0.05}$ when adjusting for multiple comparisons.}
\setlength{\tabcolsep}{5pt}
\begin{tabular}{l rrr rrr}
  \toprule
  \multicolumn{1}{c}{}   & \multicolumn{3}{c}{Bust status} & \multicolumn{3}{c}{Non-bust scores} \\
  \header{Background}     & \header{\overunder{\%}{bust}} & \header{$z$}   & \header{$p$}  & \header{\mean{score}} & \header{$t$} & \header{$p$}  \\
  \midrule
  Board             & 38   & 1.48  & 0.14 & -\$156K & 0.30  & 0.77 \\
  Finance           & 26   & -0.42 & 0.68 & -\$162K & 0.08  & 0.94 \\
  Government        &  9   & -2.05 & 0.04 & -\$107K & 2.71  & 0.03 \\
  Legal             & 35   & 0.48  & 0.63 & -\$158K & 0.21  & 0.84 \\
  Manager           & 29   & -0.04 & 0.97 & -\$150K & 0.79  & 0.44 \\
  Other             & 43   & 2.77  & 0.01 & -\$165K & 0.02  & 0.98 \\
  Physical Security & 20   & -0.49 & 0.62 & -\$146K & 0.45  & 0.68 \\
  Police            & 22   & -1.88 & 0.06 & -\$158K & 0.42  & 0.68 \\
  Risk Analyst      & 30   & 0.06  & 0.95 & -\$151K & 1.02  & 0.31 \\
  Cyber Security    & 26   & -1.10 & 0.27 & -\$173K & -0.60 & 0.55 \\
  Student           & 29   & -0.21 & 0.84 & -\$175K & -0.27 & 0.79 \\
  Technical         & 43   & 1.98  & 0.05 & -\$152K & 0.73  & 0.47 \\
  \bottomrule
\end{tabular}
\label{tab:player_bg_tests}
\end{table}

\noindent\textbf{RQ3. What patterns can be seen in the decision-making processes of players and teams?}

The previous two research questions establish which groups of participants manage to make effective risk decisions and play a better game of \ac{D-D}.  This third question explores \emph{what} decisions teams make by analyzing modal choices and clusters of decisions across game sessions, as well as investigating how the 8 teams for which we have detailed transcriptions reached these decisions. Using these, we categorize their approaches, looking for patterns and common themes between them.
Thematic analysis, a widely used method within Psychology and wider Social Sciences~\mbox{\cite{braun2006using, braun2012thematic}}, was used to explore the decision making processes of the teams. Thematic analysis lets us identify themes that ``capture[s] something important about the data in relation to the research question, and represents some level of patterned response or meaning within the dataset''\mbox{~\cite{braun2006using}}. We specifically sought to identify latent patterns that exist within the teams' discourse whilst playing \mbox{\ac{D-D}}, adhering to Braun and Clarke's~\mbox{\cite{braun2006using}} method to derive our codes and complete our analysis.
The first and fourth author completed the coding and analysis---both authors have extensive experience running \mbox{\ac{D-D}}. The two authors reviewed each transcript independently, stopping to discuss codes identified at the end of each game round they had reviewed. This ensured that the two authors developed an agreed upon codebook through their analysis (see online appendix).

\subsection{Limitations and threats to validity}
\label{sec:threats to validity}

The dataset of game sessions was collected and provided to us by the \ac{NPO}. However, in checking the data, we noticed occasional transcription errors where the recordings of the decisions made by teams did not align with their overall score in the game.  To correct these errors, we recalculated teams' scores based on the decisions recorded and the known game mechanics as the decision recordings are more likely to be accurate than the scores. 

To address the potential for the Game Master to affect the decisions made by teams during their games, Game Masters strictly follow a provided script when running games. While Game Masters were not recorded, and so cannot be explicitly modeled, the script helps to mitigate their potential influence on game decisions and outcomes.

Participant selection itself presents a threat to validity, as our dataset is captured in the wild by a third party. We have tried to counter any possible effects from this by exploring the impact that diversity of expertise has on team performance (see Section~\ref{sec:quantitative analysis}).
The dataset provided by the \mbox{\ac{NPO}} provided only a high-level description of each player's background and organization. We therefore make the assumption that a player's job description reflects their experience. Future work will include a greater indication of experience (e.g., professional qualifications held, years working) in order to be able to explore the relationship between experience and performance in a greater depth.
Similarly, we are unable to ascertain the extent to which these participants have experience of decision-making within their organizations. It is possible that some of the participants have a job where they are making high-risk decisions regularly, this in turn may provide them with an advantage. Future work should seeks to gather this additional information.

Participants may have an advantage if they play board games extensively or have played other similar exercises. All participants were asked to declare if they had played \mbox{\ac{D-D}} before they were allowed to take part in the exercise. We have run \mbox{\ac{D-D}} with thousands of players, including with professional board game developers and have found little to no advantage to players; but we cannot discount this threat.
Additionally, the participants may experience a learning effect as they play the game and learn its rules.  Expectations about how the game may be designed formed in the first few rounds (which typically took the longest to play) may influence later decision making.

Our analysis is reliant on the ability of D-D to reflect the real-world. \mbox{\ac{D-D}} is designed to be run over a short time frame and with a wide range of stakeholders. The original version of the exercise was developed and tested extensively with cyber security practitioners to reflect some of the complexity associated with cyber security decision-making within organizations. The version played in our study is based upon the experiences of the \mbox{\ac{NPO}} to reflect the problems they have encountered in organizations when responding to cyber security incidents and their experience running training nation-wide.

Our qualitative analysis is limited to a simple paired thematic analysis, to make best use of the limited qualitative data available. This is a small subset of the number of games that were played overall ($\sim$4\%), and less representative than we would like---recordings of multiple teams at adjacent tables in large sessions led to cross-talk which made an additional 13 recordings incomprehensible.
We recognize that the results must be considered exploratory, providing only a guide to the reasoning processes of decision-makers. Future studies will need to use a grounded theory-style approach and a larger sample in order to explore such reasoning processes in more detail.

\section{Impact of team make up on performance}
\label{sec:quantitative analysis}

The findings in section~\mbox{\ref{sec:findings1:impact of participant background}} address RQ1:~\emph{What effect does a player's background have on their game performance?} Section~\mbox{\ref{sec:findings1:impact of team make up on performance}}'s findings address RQ2:~\emph{What is the effect of diversity of experience on a team's performance?}

\subsection{It doesn't matter what you know\ldots}
\label{sec:findings1:impact of participant background}

The background of the individual members making up a cyber security decision-making team, as well as the sector from which their organization is drawn, are expected to have an impact on security decision-making performance. We test for such associations with our two game outcomes---whether or not the team eventually suffered a catastrophic loss and went \emph{bust}, and for teams which did not go bust, their final score in terms of financial penalties suffered.

\subsubsection{Player background}

Our primary interest is whether any subgroups of teams by player background (Table~\ref{tab:skills}) are significantly different from the overall population values for bust status and average non-bust score. Overall, 30\% of teams go bust, and the average score for non-bust teams is -\$165,020.50. To identify if any subgroups deviate significantly from this, and correcting for multiple comparisons with the Bonferroni method, a series of
two-sided one-sample z-tests (for bust status) and t-tests (for average score) were performed (Table~\ref{tab:player_bg_tests}). No such groupings were found to differ significantly from the population averages. The lack of significant effects for score outcome were also observed in both game version sub-populations. These results suggest that no group of teams, by player background, differs from the population than might be expected by chance. A similarly-corrected pairwise analysis between all subgroups showed no significant differences in bust status or average non-bust scores.

As mixed team composition could mask the effect of player background, we also studied homogeneous teams of one type of experts separately (and a combined category for heterogeneous teams).  The results, shown in Table~\ref{tab:player_bg_homo_tests}, also find a null result for player background within homogeneous teams. Pairwise analysis also finds a null result between all groups. Bearing in mind that group differences are still best explained by chance variation from common population mean, we see that the largest subset (cyber security specialists) performed worse than average in terms of non-bust scores, but also went bust less often.  \emph{This may suggest that cyber security specialists try to avoid disaster, at the cost of suffering smaller, less damaging, attacks more often.}

\begin{table}[t]
  \tablestyle
\caption{Breakdown of game outcomes for homogeneous teams. No tested subgroups were significantly different from the overall means at $\mathbf{\alpha=0.05}$. Tests were inappropriate for the smallest subgroups.}
\setlength{\tabcolsep}{5pt}
\begin{tabular}{lr rrr rrr}
  \toprule
    \multicolumn{2}{c}{} & \multicolumn{3}{c}{Bust status} & \multicolumn{3}{c}{Non-bust scores} \\
  \header{Background}     & \header{n}   & \header{\overunder{\%}{bust}} & \header{$z$}   & \header{$p$} & \header{\mean{score}} & \header{$t$}   & \header{$p$}  \\
  \midrule
  Heterogeneous  & 103 & 32  & 0.49  & 0.62 & -\$156K & 0.84  & 0.41 \\
  Cyber Security & 35  & 20  & -1.27 & 0.21 & -\$174K & -0.43 & 0.67 \\
  Police         & 26  & 23  & -0.75 & 0.45 & -\$170K & -0.23 & 0.82 \\
  Student        & 21  & 28  & -0.12 & 0.90 & -\$175K & -0.27 & 0.79 \\
  Other          & 11  & 45  & 1.13  & 0.26 & -\$191K & -0.91 & 0.41 \\
  Risk Analyst   & 6   & 33  & 0.19  & 0.85 & -\$141K & 0.52  & 0.64 \\
  Manager        & 2   & 50  &       &      & -\$272K &       &      \\
  Government     & 2   & 0   &       &      & -\$155K &       &      \\
  Technical      & 1   & 100 &       &      &         &       &      \\
  Board Member   & 1   & 100 &       &      &         &       &      \\
  \bottomrule
\end{tabular}
\label{tab:player_bg_homo_tests}
\end{table}

Given that individuals from different backgrounds may be unevenly distributed across teams, the above averages do not adequately explain the contribution that a member of each background makes to the game outcomes. To determine this weighting, we fit two models using team composition information. First, a multiple regression was fit for the average non-bust score, using team composition and including the game version as a control for the previously-described effect this has on scores (with the post-\ac{GDPR} game version taken as the reference level, since this represents the majority of games). Second, a logistic regression was fit to predict bust status from team composition information only. The results are shown in Table~\ref{tab:player_bg_coeffs}.
Both models are not significant and a poor fit ($R^2=\{0.038,0.065\}$), and as such these coefficients must be interpreted cautiously, as they may not generalize.

Each coefficient can be understood as the effect on the overall team outcome produced by including a single member with the corresponding background in the team, relative to the values of the intercept and game version.
The background with the greatest positive effect on odds of going bust (i.e., increasing odds of going bust) was Technical, followed by Board Member and Other. The backgrounds with the greatest negative effect on odds of going bust (i.e., reducing odds of going bust) were Physical Security and Government. The backgrounds with the most deleterious effect on the final score of non-bust teams were Finance and Legal, and those with the most positive effect were Government and Technical. In line with previous discussion, cyber security specialists have a moderate negative association with the expected non-bust score.

\begin{table}
\tablestyle
\caption{Regressed coefficients for a unit increase in team composition, in terms of non-bust scores and log odds of the team going bust.}
\begin{tabular}{l rr rr}
  \toprule
  \multicolumn{1}{c}{} & \multicolumn{2}{c}{Bust status} & \multicolumn{2}{c}{Non-bust scores} \\
  \multicolumn{1}{c}{} & \multicolumn{2}{c}{(logistic)} & \multicolumn{2}{c}{(linear)} \\
  \multicolumn{1}{c}{} & \header{log odds} & \header{z value} & \header{$\beta$}   & \header{t value} \\
  \midrule
  (Intercept)       & -1.27 & -1.59 & -103,961.28 & -2.01 \\
  Version 1         &       &       & -47,901.37  & -2.16 \\
  \addlinespace
  Board             & 0.27  & 0.94  & -17,231.96  & -0.96 \\
  Finance           & -0.15 & -0.39 & -27,853.44  & -1.37 \\
  Government        & -0.57 & -0.91 & 4,997.56    & 0.28  \\
  Legal             & 0.17  & 0.37  & -16,216.28  & -0.62 \\
  Manager           & 0.03  & 0.11  & -9,540.74   & -0.67 \\
  Other             & 0.26  & 1.23  & -10,999.24  & -0.81 \\
  Physical Security & -0.58 & -0.49 & -11,453.10  & -0.20 \\
  Police            & -0.01 & -0.05 & -4,698.13   & -0.38 \\
  Risk Analyst      & 0.09  & 0.40  & -5,782.28   & -0.42 \\
  Cyber Security    & 0.04  & 0.26  & -11,354.32  & -1.12 \\
  Student           & 0.08  & 0.40  & -8,191.02   & -0.73 \\
  Technical         & 0.43  & 1.43  & -9,679.68   & -0.41 \\
  \bottomrule
\end{tabular}
\label{tab:player_bg_coeffs}
\end{table}

\subsubsection{Organization sector}

We used a series of Bonferroni-corrected two-sided one-sample z-tests and t-tests (Table~\ref{tab:org_bg_tests}) to identify any significant deviations in either game outcome from the population averages among teams as grouped by organizational background (Table~\ref{tab:backgrounds}).
We found no significant difference from the population averages for any organizational subgroup---implying that organization background is also unrelated to performance at \ac{D-D}.
The lack of significant effects for the score outcome are also observed in the \emph{pre-\ac{GDPR}} game version sub-population, but within the \emph{post-\ac{GDPR}} version sub-population, the average score of -\$60,500 for the 8 Academia teams was found to be significantly higher than the \ac{GDPR} sub-population mean (Cohen's $d=0.95, t=8.23, p=0.0012$). These teams were all students participating in a program for gifted security students from a particular single student challenge game session (and represent all teams from this session), suggesting a common explanation unrelated to the organization sector.

\begin{table}
  \tablestyle
\centering
\caption{Breakdown of game outcomes by organizational background of teams, with test statistics and uncorrected p-values. No subgroups were significantly different from the overall means at $\alpha=0.05$ when adjusting for multiple comparisons. Tests were not appropriate for the \ac{SME} subgroup (3 teams).}
\begin{tabular}{l rrr rrr}
  \toprule
  \multicolumn{1}{c}{} & \multicolumn{3}{c}{Bust status} & \multicolumn{3}{c}{Non-bust scores} \\
  \header{Sector}           & \header{\overunder{\%}{bust}} & \header{$z$}   & \header{$p$}  & \header{\mean{score}} & \header{$t$}   & \header{$p$}  \\
  \midrule
  Private          & 35 & 1.01  & 0.31 & -\$157K & 0.52  & 0.61 \\
  Police           & 19 & -1.62 & 0.11 & -\$177K & -0.73 & 0.47 \\
  Public           & 31 & 0.14  & 0.89 & -\$169K & -0.23 & 0.82 \\
  Mixed            & 35 & 0.52  & 0.60 & -\$172K & -0.27 & 0.79 \\
  Academia         & 32 & 0.17  & 0.87 & -\$167K & -0.05 & 0.96 \\
  Business Support & 10 & -1.37 & 0.17 & -\$143K & 0.84  & 0.43 \\
  Unknown          & 40 & 0.50  & 0.62 & -\$209K & -1.74 & 0.22 \\
  \ac{SME}         & 33 &       &      &  -\$95K &       &      \\
  \bottomrule
\end{tabular}
\label{tab:org_bg_tests}
\end{table}

\subsection{Diversity of expertise doesn't help\ldots}
\label{sec:findings1:impact of team make up on performance}

Beyond the specific background of team members, a general diversity of viewpoints is often considered an asset to decision-making~\cite{simons1999making,olson2007strategic}. But what effect does diversity of expertise have on \ac{D-D} performance?

Comparing the bust rate and average non-bust scores, we find little to separate homogeneous and heterogeneous teams.
For the bust rate, a one-sided test for equality of proportions between bust counts in homogeneous and heterogeneous teams fails to reject the null that the proportion is the same ($p=0.29$, 28\% and 32\%). To compare whether there is a significant difference between average scores in the homogeneous and heterogeneous groups we use a Welch one-sided two-sample t-test.  We fail to reject the null that the average scores are the same between these groups ($p=0.14$).
The lack of significant effect for the score outcome is also observed in both game version sub-populations.

Does increasing the diversity of expertise in a team improve game outcomes? It is after all possible that many heterogeneous teams may simply not be heterogeneous \emph{enough} to show an effect.
Using the Shannon index as a measure of diversity~\cite{spellerberg2003tribute}, we find that neither game outcome is statistically significantly associated with rising diversity.
For the bust state, a Welch one-sided two-sample t-test between the Shannon index values for the bust and non-bust teams fails to reject the null hypothesis ($p=0.63$).
For the non-bust scores, we measure association between the diversity (Shannon index) and teams' final scores by calculating Pearson's product moment correlation between them. The results show that, while the overall correlation is positive ($r=0.11$), it is small and not statistically significant ($p=0.09$).

The lack of significant effect for the score outcome is also observed in both game version sub-populations. These results together suggest that increased diversity of expertise (player background) has no strong effect on team performance.
Diversity of opinions, it would appear, doesn't help teams play a better game.

In summary, neither player background, organization sector, nor makeup of the team have a significant impact on the relative performance of teams. One interpretation of these findings is that \mbox{\ac{D-D}} is readily accessible---prior experience offers no particular gain, making it a potentially valuable training tool. It may also indicate that expertise is not always necessary for making ``valid'' cyber security decisions, but it is important to note that these findings only hold within the context of \mbox{\ac{D-D}}.

\section{How do teams make their choices?}
\label{sec:qualitative analysis}

This section addresses RQ3:~\emph{What patterns can be seen in the decision-making processes of players and teams?}

Having established that neither team background nor organization type has a differentiating impact, we explore the specific decisions made by the 8 teams for which we have transcripts.  This analysis was completed by two of the authors and took the form of thematic analysis~\mbox{\cite{braun2006using,braun2012thematic}}. This involved the two authors reading through each transcript together. Each author made annotations and notes as they read through. At the end of each round the reviewers stopped to discuss their observations and identified key common themes to carry forward.

The 8 teams which were reviewed happened to be homogeneous teams of security experts (see Table~\ref{tab:qual_teams}). Teams T3 and T8 both went bust during their games. T2 had the greatest losses recorded out of all 208 teams without going bust. Teams T1, T5, T6 and T7 also scored better than the average score of -\$174,053 (for teams that did not go bust and were comprised entirely of cyber security specialists, see Table~\ref{tab:player_bg_homo_tests}). T4 performed slightly worse than the average overall. It is worth noting that T3 were the largest single team that played the game, and that despite the size of the group, all participants were still able to contribute to the conversation.

\begin{table}
  \tablestyle
  \caption{Teams analyzed qualitatively.}
  \begin{tabular}{rrrr}
    \toprule
    \header{Team} & \header{Size} & \header{Final Score} & \header{Duration (mins)} \\
    \midrule
    T1 & 6  & -\$150,000 & 90 \\
    T2 & 6  & -\$570,000 & 45 \\
    T3 & 10 & -\$1,265,000 & 88 \\
    T4 & 4  & -\$205,000 & 78 \\
    T5 & 4  & -\$90,000 & 73 \\
    T6 & 4  & -\$95,000 & 44 \\
    T7 & 5  & -\$125,000 & 46 \\
    T8 & 4  & -\$1,140,000 & 77 \\
    \bottomrule
  \end{tabular}
  \label{tab:qual_teams}
\end{table}

\begin{figure}
    \centering
    \includegraphics[width=0.35\linewidth,angle=90]{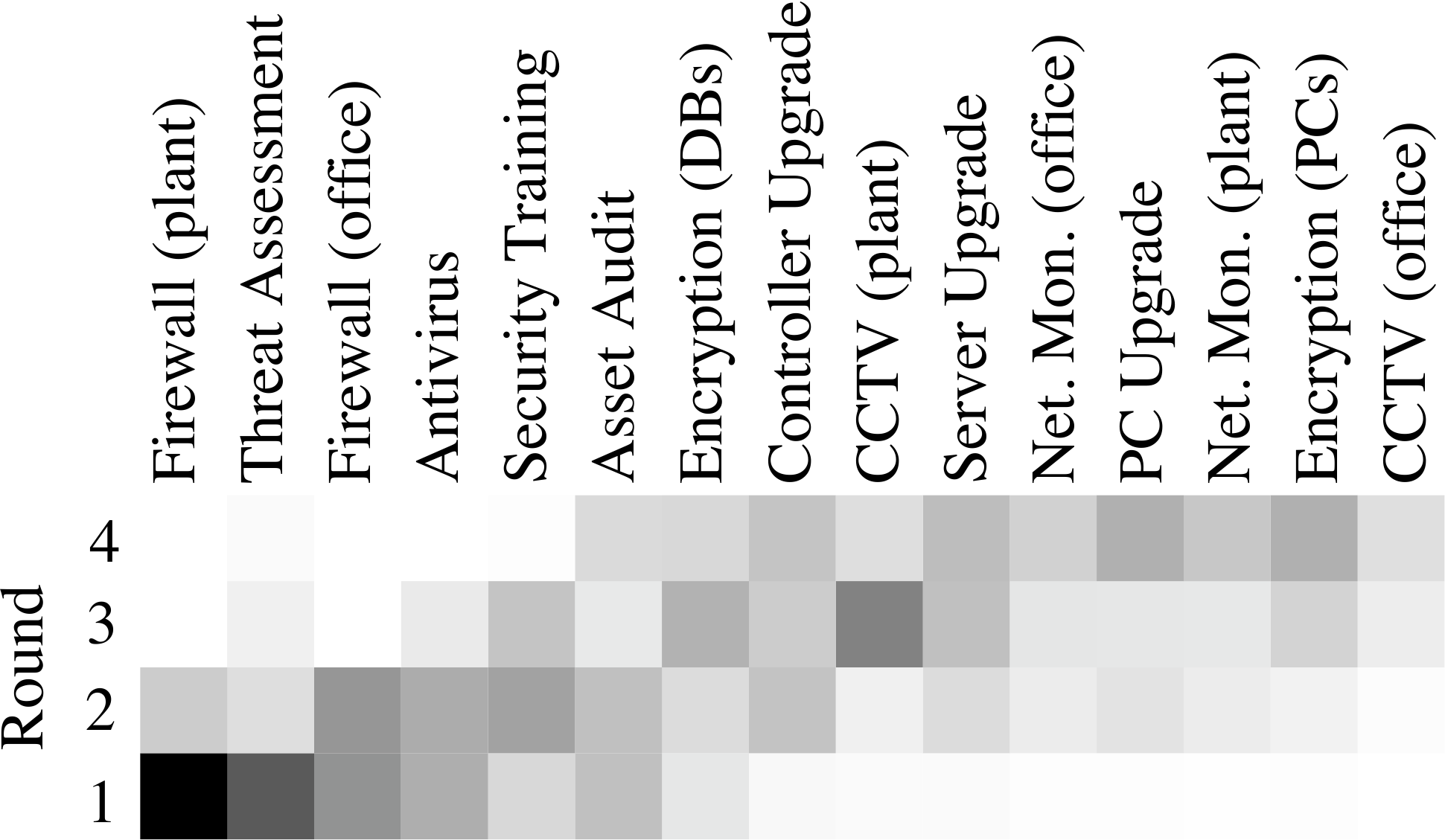}
    \caption{Heatmap of investments made over game rounds by all 208 teams.  The darker the shade, the more often the investment was played during that round.}
    \label{fig:rounds}
\end{figure}

\subsection{Why did you buy that?}
\label{sec:findings2:why did you buy that?}

Our analysis reveals a number of general decision-making patterns when teams invest in defenses.

\subsubsection{Start with what you know or what you've heard of\ldots} 
\begin{figure}
    \centering
    \includegraphics[width=0.8\linewidth]{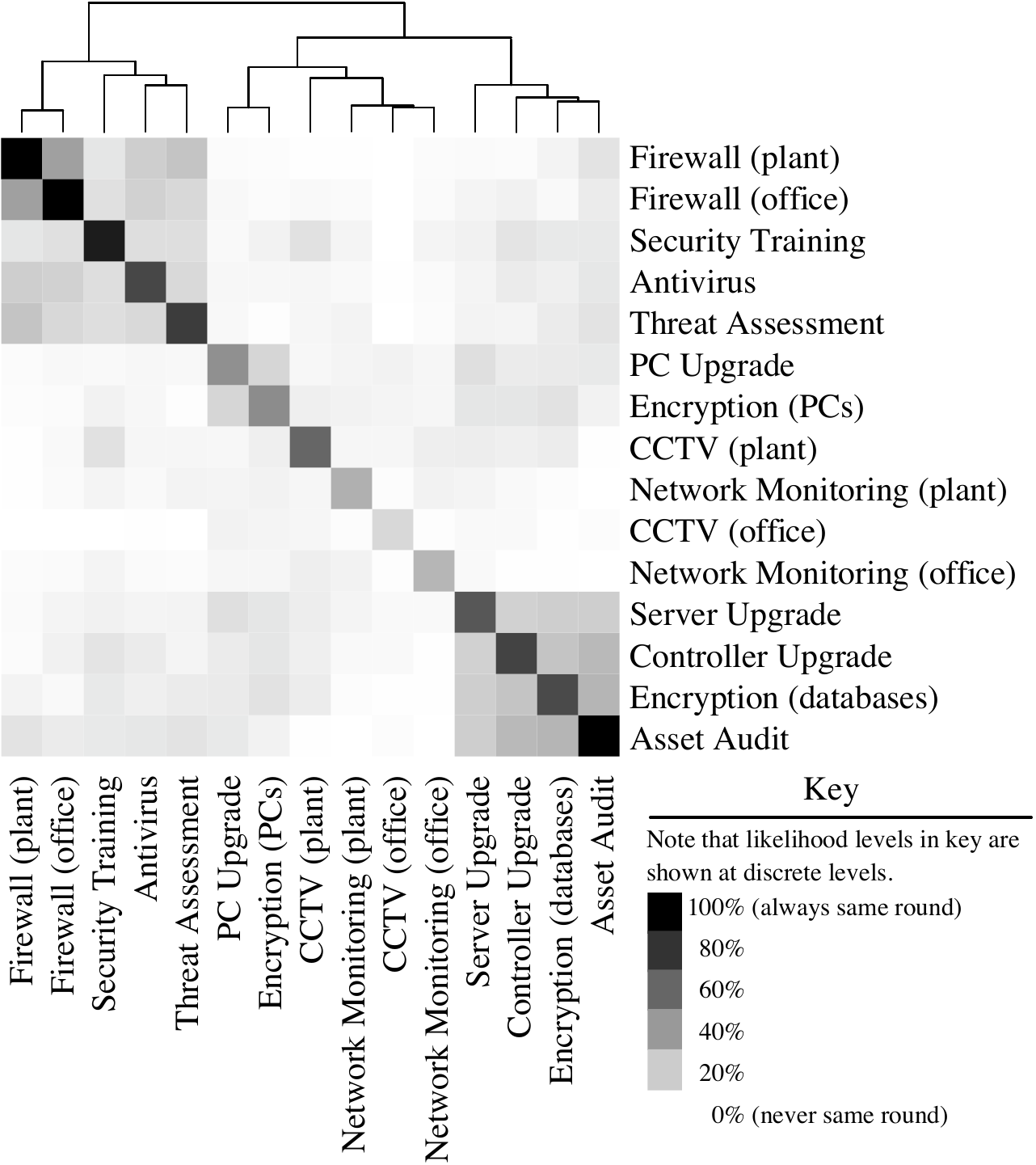}
    \caption{Decisions of all 208 teams plotted by the number of times they co-occur. Hierarchical clustering by the Ward minimum variance method.}
    \label{fig:cooccurrences}
\end{figure}

Figure~\ref{fig:rounds} shows that all 208 teams tend to focus on similar investments in Round~1: \emph{Firewall (Plant), Threat Assessment, Firewall (Office), Antivirus, Security Training} and \emph{Asset Audit}. Figure~\ref{fig:cooccurrences} shows the co-occurrence of investments (the chance that one investment was played in the same round as any other) and three distinct clusters: one around the \emph{Firewall (Plant)}, a second around the \emph{PC Upgrade} and \emph{Encryption}, and a third around the \emph{Asset Audit} and investments it enables. The cluster around \emph{Firewall (Plant)} indicates that most of the 208 teams that invested in \emph{Firewall (Plant)} during Round 1 also invested in a combination of \emph{Firewall (Office)}, \emph{Security Training, Antivirus} and the \emph{Threat Assessment}. Teams were also likely to invest in the \emph{Asset Audit}, but to a lesser extent.

In our qualitative analysis, we found that teams tended to spend a lot more time talking during this opening round than in any of the others. However, during this extensive dialogue, teams rarely dedicated much time to discussing the benefits of their range of investment options. For the most part, teams appeared to review the opening options before, very quickly, identifying the six \emph{``obvious''} candidates. For example, T2 remarked when considering the \emph{Firewall}:

\begin{ddquote}
    ``But you've got to have a Firewall, haven't you?''
\end{ddquote}

\noindent As Figure~\ref{fig:rounds} indicates, the majority of the 208 teams made similar choices during Round~1. All the transcripts we have reviewed are from homogeneous teams of cyber security experts, however that the early choice of a \emph{Firewall} is reflected in the wider dataset suggests that people have learned about the common forms of security mechanisms.

\subsubsection{Thinking is a luxury}

The 8 teams demonstrate reasoning from simplistic to more justified, however, some teams also develop flawed assumptions and become confused.
Teams often used very simplistic reasoning for their choices, particularly during the first round of the game. For example, most of the 8 teams prioritized investing in \emph{Firewalls} without ever considering whether a \emph{Firewall} would actually help to defend their system.  When discussing whether they should opt for the \emph{Asset Audit} and \emph{Threat Assessment} or the two \emph{Firewalls} T1 noted:

\begin{ddquote}
  ``Let's say I was the \emph{[operator]} and you were on to take in this \emph{[Threat]} Assessment and Asset Audit and during the first quarter and and that the shit hits the fan and you then the \emph{[boss]} would look and say, well she had no Firewalls and you know\ldots''
\end{ddquote}

\noindent These preferences are echoed in other teams' games, with participants noting that \emph{``you have to have a Firewall''}, without ever stating \emph{why} they needed it.  There is often debate about whether a mitigation is required or not, but there is little---if any---debate about the \emph{Firewalls}.
However, that isn't to say that teams neglect to use reasoning at all during the first round. Some teams spent more time developing more substantial and justified reasoning. For example, T3 dedicated time to ensuring they evaluated their potential assets.

\begin{ddquote}
    ``if you don't know what your assets are, we're not just talking physical assets. Information assets come into it, various other things. Even people are assets. You've got to think about what you need to protect and how you are going to protect them. You can't do that if you don't understand what they are.''
\end{ddquote}

\noindent This difference in depth of reasoning often varies, with some teams demonstrating minimal reasoning throughout the game and others demonstrating a lot. Some teams have a tendency to employ reasoning around specific actions or situations. For example, T6 dedicated much time to identifying whether the Plant or Office site should be the focus of their investments based on the relative complexity of rebuilding either:

\begin{ddquote}
    ``I can probably recover the office environment. I can move that \emph{[office]} elsewhere much faster than I can move this \emph{[plant]}.''
\end{ddquote}

\noindent Meanwhile, other teams demonstrate confused reasoning; developing ideas and assumptions that are inherently flawed. This was particularly manifest early on in the game when teams were discussing the need for \emph{Firewalls}. This is especially interesting as the teams in our qualitative dataset are all security experts. For example, T7 suggested that having a \emph{Firewall} might actually make the company a target:

\begin{ddquote}%
\emph{Speaker 4:} ``\ldots{}let's say you've got 10 companies and one's got a Firewall, you've gotta find which Firewall it is, particularly a good Firewall, then that's the company they're gonna target.''

  \emph{Speaker 5:} ``\emph{(Laughing)} Are they gonna target the one with the good Firewall?''

  \emph{Speaker 4:} ``Yeah.''

  \emph{Speaker 5:} ``Under the assumption that---''

  \emph{Speaker 3:} ``There could be something good in there.''
\end{ddquote}

\subsubsection{Shiny is always better}

Several of the teams explored the possibility that investment in advanced technology would negate the need to cover security basics like \emph{Firewalls, Antivirus, Security Training} and \emph{Software Patching}, following the mantra:

\begin{ddquote}[T1]
``It's always good to have new shiny new equipment''
\end{ddquote}

\noindent T2 proposed buying new security tools ahead of upgrading existing systems, noting that:

\begin{ddquote}
\emph{Speaker 3:} ``We don't need to upgrade. You can put a further level of\ldots''

\emph{Speaker 6:} ``\ldots{}Anti-virus\ldots''

\emph{Speaker 3:} ``Yeah you can put a further level of security rather than upgrade the assets.''
\end{ddquote}

\noindent T4 meanwhile actually train organizations in the importance of cyber security and yet they pursued a similar course while simultaneously acknowledging that neglecting the basics was completely contrary to their own training:

\begin{ddquote}
    ``If we don't take that [Antivirus], we're actually going against everything we say in our own presentation.''
\end{ddquote}

\noindent We observed several other instances where teams debated whether the (high-tech) \emph{Network Monitoring} investment could be used in lieu of both the \emph{Firewalls, Antivirus} and \emph{Security Training}. They believed that the advanced nature of the technology would negate the need to invest in the basics. Trusting technology to provide security is a known user strategy~\cite{dourish2004security,wash2015too}, though it is interesting to contrast expert and non-expert opinions.  In a study on user behaviors to stay safe online Ion~et~al{.} found that non-experts favored using technical products, such as anti-virus, to provide security but dismissed the value of updates~\cite{ion2015no}. In contrast, the experts believed the opposite, having a relatively low opinion of products and a strong belief in the value of updates~\cite{ion2015no}. Our dataset (both quantitative and qualitative) suggests that cyber security experts are just as prone to this type of bias. 

\subsubsection{The best laid plans\ldots{}or lack thereof\ldots{}}

Threat modeling is commonly used by security professionals to help identify and evaluate threats and is often considered a key aspect of best practice, forming an important part of both ISO27001~\cite{iso27001} and NIST SP 800-53~\cite{nist-800-53} standards. Only T8 actually demonstrated the use of a threat model of some kind. Interestingly, T8 played the game at a major cyber security event and referred back to a presentation that they had encountered earlier that day introducing four factors of risk to SMEs (including Phishing, Insider Threats, DDOS and Ransomware):

\begin{ddquote}
    ``So you've got phishing, you've got ransomware, and without Antivirus you've got zero protection for that. DDOS, you've got zero protection if you don't have erm\ldots{}Firewalls can do some of that, erm\ldots{}and the other option for it is, isn't on there. So you've got, we've gotta work out whether we're worried more about insider threats or outsider threats.''
\end{ddquote}

\noindent Later on in the game the team reflect on this particular set of threats and realize that they were in fact meant specifically for \acp{SME}. They, therefore, decide to refine their threat model:

\begin{ddquote}
    ``What I'm thinking is we were going from the four key factors for \acp{SME}, and actually we've gotta think bigger here because this is electricity, electricity is of interest to \emph{[nation]} states.''
\end{ddquote}

\subsubsection{Forward planning}

We found that some of the decisions that teams make are influenced by their plans for the future. The investments available in D-D each have a cost associated with them (see Table~\ref{tab:investment-costs}). When playing the game participants have a finite budget to spend in each round---with any remainder carrying over into the next round. When analyzing the choices all 208 teams made on a round by round basis, we found a non-significant but suggestive correlation ($p=0.06$) that the cheaper options were favored over the more expensive choices (correlation between cost and the number of times the investment was purchased $r =-0.67$). We also observed trends in the way that teams evaluated certain items during the first round. Most teams in our 8 transcripts, for example, quickly decided that the \emph{CCTV} could be put off until a later stage in the game.

\begin{ddquote}[T5]
    ``This is year one. Can't afford to be putting CCTV in yet. That's like\ldots{}a luxury at the moment for this company.''
\end{ddquote}

\subsubsection{The Cost of Investment}
We also found a suggestive but non-significant correlation ($p=0.09$) between the cost of the investment and the round in which the investment was purchased on average ($r = 0.45$). While these relationships are not significant, they suggest that there may be a small bias towards choosing more and cheaper security investments over any larger more expensive investments until a pressing need is demonstrated.  At the end of Round 2 in the game, if a team has not purchased the \emph{CCTV (Plant)} investment, the players suffer a physical intrusion into the plant.  This encourages players to invest in the CCTV in Round 3, however the suggestive correlation is still seen even if we ignore the \emph{CCTV (Plant)} ($r=0.43$, $p=0.12$).

One explanation could be that teams choose one investment as a primary defense in a round and then use the cheaper options to \emph{use up} any extra budget they have at the end of a round.  While in \ac{D-D} unused budget is rolled over to the next round, teams appear to like to spend any excess capital they have in the current round to get extra protection there and then, rather than hoard it:
\begin{ddquote}[T2]
  ``If we want the anti-virus, we've got 40,000 left.  Then you could buy a couple of 20Ks if you wanted?  What could we get?  PC Encryption, data Encryption, uh, or a Threat Assessment\ldots''
\end{ddquote}

\noindent Other teams like to try and plan their spend over multiple rounds before implementing their plan.

\begin{ddquote}[T5]
``\ldots we need to be thinking about year three and four.''
\end{ddquote}

\subsubsection{Fixation on Single Issues}

In several instances teams fixated on one particular vulnerability, threat actor or investment. This serves to confirm the notion of \emph{tunnel vision} initially observed by Frey~et~al{.}~\cite{Frey2019TheGoodBadandUgly}. For example, T8 spent a lot of time considering the fact that the two databases were replicated. In particular they were concerned that the data was both transmitted between databases and stored in an unencrypted format:

\begin{ddquote}
    ``I get why they've done this, because they're replicating their, their two databases there which is fine, so they've got an off-site copy of it. But it's going over the public internet and there is no security, perimeter security at all, to be able to go through, to be able to stop any type of attack or any type of compromise that's happened on a dataset that's going across the public internet.''
\end{ddquote}

\noindent This concern was raised before they had played the \emph{Asset Audit} and the \emph{Database Encryption} investment was available. While the ability to identify this vulnerability demonstrates experience, it didn't necessarily help the team. They fixated on this issue and neglected to consider vulnerabilities within the other systems, 
causing their organization to go bankrupt (a possible outcome in \mbox{\ac{D-D}} if teams suffer a large or repeated breaches---a 2019 survey found that 25\% of \mbox{\acp{SME}} went bankrupt after suffering a breach~\mbox{\cite{ncsa2019small}}).

Section~\mbox{\ref{sec:findings2:why did you buy that?}} outlines the common investment decision-making trends exhibited by teams. Practitioners should pay particular attention to the impact that these biases can have on decision-making. For example, our analysis demonstrates how decision-makers are more likely to value those investments with which they have greater familiarity. Likewise, decision-makers are also likely to favor cheaper investments over optimal investments---this may explain why teams deferred investing in CCTV until the later rounds of the game, despite acknowledging the importance of physical security. How teams treat the notion of physical security is another potentially worrying bias---during the game teams often treat physical and cyber security as mutually exclusive. Within the context of \mbox{\ac{D-D}} these biases can have a major impact on the way that decisions are made (e.g., the depth to which decisions are explored) and on the final choices made. Organizations should therefore set out to raise awareness of these biases in their decision-makers in order to negate their impact wherever possible.

\subsection{You can't handle the truth}
\label{sec:findings2:You can't handle the truth}

The ability to learn new information and refine hypotheses is key to decision-making. During the game, teams gain new information at the end of each round when they suffer attacks which relate to their investment choices. The game also provides teams with two additional information gathering cards: \emph{Threat Assessment} and \emph{Asset Audit}. Yet teams often displayed an \emph{unwillingness to learn} from the new information they gained or could gain from these.

\subsubsection{This Threat Assessment is rubbish\ldots{}}

Teams often play the \emph{Threat Assessment} card before the \emph{Asset Audit} card (see Figure~\ref{fig:rounds}), typically playing it in the first round. The card provides teams with a simple summary of the three types of attacker they are likely to encounter (Script Kiddies, Organized Crime Groups and Nation State) and common attack vectors used by each. However, several teams in our transcripts of security experts were dismissive of the new information, while falling victims to the very attack vectors it highlighted.

\begin{ddquote}[T3]
     ``Well, the Threat Assessment was not useful at all, and that should never have been paid for\ldots''
\end{ddquote}

\subsubsection{Responses to learning about vulnerabilities}

\begin{figure}
  \centering
  \includegraphics[width=0.8\linewidth]{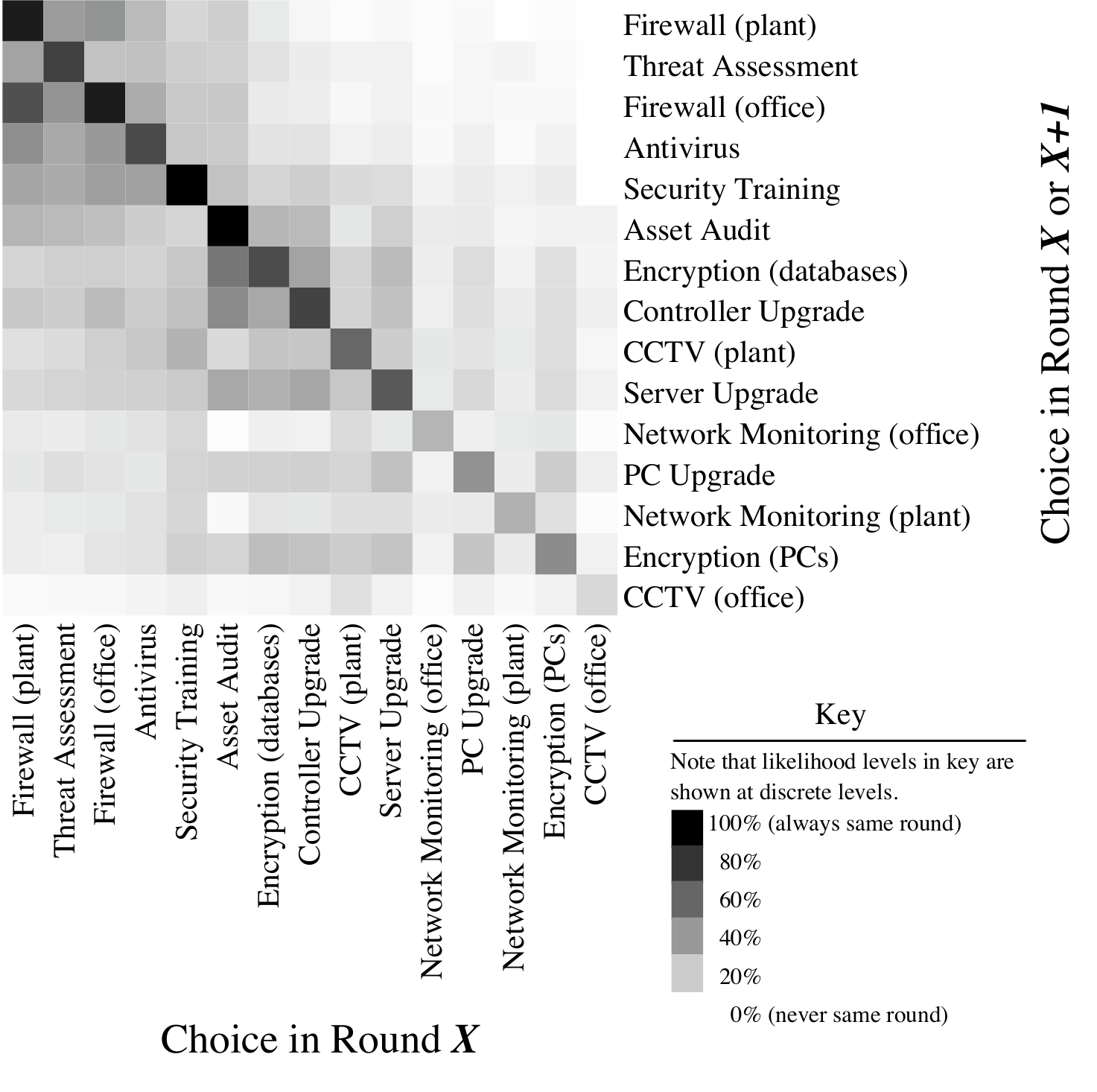}
  \caption{
  Investment decisions (horizontal) of all 208 teams plotted for the number of times they co-occur with or immediately precede other decisions (vertical). }
  \label{fig:same-round-or-next}
\end{figure}

Investing in the \emph{Asset Audit} provides teams not only with information about their vulnerabilities but it also unlocks five additional defense investments (\emph{PC Upgrade, Server Upgrade, Controller Upgrade, PC Encryption, Database Encryption}). For most teams it serves as a catalyst, often justifying earlier parts of their conversation where they might have speculated about the state of their fictional organization. Figure~\ref{fig:same-round-or-next} shows the number of times each investment (horizontal axis) appears in the same round as or immediately before a second investment (vertical axis).  Teams that chose the \emph{Asset Audit} investment commonly play the \emph{Database Encryption, Server Upgrade} and \emph{Controller Upgrade} in the same or subsequent rounds. This may appear an obvious consequence of the game mechanics, but it is worth observing that it does not apply equally across all unlocked options. The \emph{Asset Audit} also reveals the \emph{PC Upgrade} and \emph{PC Encryption} cards, yet these tend to get played together (Figure~\ref{fig:cooccurrences}) later on in the game and are rarely considered an immediate priority.
However, despite the benefit which the \emph{Asset Audit} affords, some of the 8 teams in our transcripts remained skeptical:
\begin{ddquote}[T2]
  ``I feel as like I'm being fleeced by the Asset Audit company that fixed my WIFI for free and is trying to get me to buy all this new kit.''
\end{ddquote}

\subsubsection{A Threat Assessment is going to tell me about my vulnerabilities right? RIGHT?!}

Our qualitative analysis leads us to question whether teams see the difference between threats and vulnerabilities. The game is designed such that if played, the \emph{Threat Assessment} will provide teams with a list of threat actors that they may encounter and how they may attack. A description to this end is included on the card itself and the Game Masters specifically mention to the teams that they should read these descriptions. The \emph{Asset Audit} meanwhile provides an overview of the vulnerabilities within the current system, and
results in the teams receiving additional cards.

We encountered several instances where teams spent a lot of time clearly confused as to the difference between threats and vulnerabilities. This was unexpected given that the teams were all practicing cyber security specialists. Even after teams had read the card descriptions, they still expected the \emph{Threat Assessment} to tell them about their vulnerabilities. This misunderstanding is something the authors have seen while observing other instances of \ac{D-D} games run by the \ac{NPO}. We would postulate that this confusion explains why so many teams play the \emph{Threat Assessment} before the \emph{Asset Audit}---what they really want to know about are the organization's vulnerabilities (see Figure~\ref{fig:rounds}).

\begin{ddquote}[T3]
``We, we've always said around the table, several times now, we got a Threat Assessment \emph{[in Round 1]}, we didn't get a vulnerability assessment. That's really what we needed. We assumed\ldots''
\end{ddquote}

\noindent Other teams recognize the difference but still debate which of the two intelligence cards should be played first:

\begin{ddquote}
     ``Why do the threats first? Why not do the Asset Audit first so that you know what needs protecting the most''
\end{ddquote}


\subsection{Nobody expects the unexpected}
\label{sec:findings2:nobody expects the unexpected}

The attacks suffered and financial ramifications affect the decisions that teams make. The majority of teams encounter the same attack, a physical break-in at the plant at the end of Round~2 (if they haven't purchased the \emph{CCTV} by this point). This results in a knee-jerk response with a large number of teams investing in \emph{CCTV} for the plant in Round~3 (see Figure~\ref{fig:rounds}). Not all teams respond in this way, for example, T7 just brush the attack to one side---they never install \emph{CCTV} and as a result fail to spot someone retrieving a keylogger left during the break-in, the resultant data breach bankrupts their organization:

\begin{ddquote}
``If we had CCTV we'd know!''
\end{ddquote}

\noindent The way that teams respond to attacks varies from team to team, some are highly reactive to attacks, others meanwhile attempt to rationalize the attacks that they have suffered. For example, T5 chose not to invest in \emph{Antivirus} during the first round and they consequently suffered a ransomware attack, meaning that they had to pay to unlock their PCs. However, despite this, the team still did not think that the \emph{Antivirus} was a worthwhile investment during the second round. Instead they explored whether \emph{Security Training} could address the same issue:

\begin{ddquote}
 ``The training comes in so once you get against that threat, that's swept out.''
\end{ddquote}

\noindent The same team then suffered a further malware attack at the end of Round 2, and they reluctantly acknowledged that malware attacks were a cause for concern:

\begin{ddquote}
    ``\ldots I think that's the key one \emph{[Antivirus]}, because it seems to be year 1, if we had played it, it might have stopped something\ldots{}year 2, that was the key thing. Everything. All the losses, reputations, from that. So that needs playing.''
\end{ddquote}

\noindent
Sections \mbox{\ref{sec:findings2:You can't handle the truth}} and \mbox{\ref{sec:findings2:nobody expects the unexpected}} both explore how teams respond to receiving new information and evaluating its value.  In \mbox{\ac{D-D}}, teams often conflate threats and vulnerabilities. Organizations should encourage decision makers to think more carefully and distinctly about problems in the systems they own, processes they use and services they purchase (vulnerabilities), and from external threats that may be interested in targeting them.
Our analysis also suggests that decision-makers often react to adverse information in one of two ways: an immediate knee-jerk reaction, or a more laissez-faire response. Both are problematic, with knee-jerk responses in the game often seeking to address the symptoms only, neglecting the underlying cause. More laissez-faire approaches often assume that attacks are mutually exclusive, leaving them open to repeat attacks. Organizations and practitioners should therefore set out to try and take measured responses to incidents.

\section{Related Work}
\label{sec:related-work}

Recent work by Stevens~et~al{.}~\cite{Stevens-2018-The-battle-for-new-york} suggests that training in risk analysis methods can increase risk identification. They trained 25 members of the New~York City Cyber~Command to use the Center~of~Gravity framework~\cite{eikmeier2004center}; a risk approach.
They found that 20~of~25 participants had incorporated threat modeling into their work and made some significant security improvements as a result. Our analysis highlights critical need for such training as, among 208 teams, we found only one instance where the team applied a formal risk approach. 

Cyber security decision-making does not, however, happen in isolation\textemdash with the majority of responses to risk requiring an investment of some kind. To this end, the majority of organizations are concerned with weighing up the cost of cyber security against perceived risk. \emph{Investment related} research therefore helps organizations to identify priorities by attempting to capture the potential monetary consequences of their actions \cite{cavusoglu2004model,gordon2002return,sonnenreich2006return,fielder2016decision,gordon2015externalities,gordon2016investing}. This ultimately results in metrics such as the \emph{Return on Security Investment (ROSI)} approach~\cite{cavusoglu2004model,gordon2002return,sonnenreich2006return}. Such methods attempt to offer a practicable middle ground for organizations, encouraging them to consider the impact of their security decisions while acknowledging that they are ultimately profit-driven. This tension is captured in the design of the D-D game that teams encounter; teams have to identify which choices to make while working within a finite budget. Teams also encounter financial ramifications for their choices, all of which serve to replicate this tension.

M'manga~et~al{.}~\cite{mmanga2017folk}, suggest that risk analysis is rarely a rational process and risk interpretation is influenced by many factors:
\emph{Awareness, Communication, Tool Capabilities} and \emph{Individual Capabilities}.  Jalali~et~al{.}~\cite{jalali2019decision} used a simulation game to compare the performance of 38 security professionals with a control group consisting of 29 postgraduate students. They found that experienced groups had no better understanding than the non-experts that it takes time to establish cyber security capabilities. Likewise, both groups made similar errors when dealing with the uncertainty of cyber security incidents. Their analysis is based on individuals making decisions based on the simulations, whereas our dataset reflects team decision-making. Furthermore, Jalali et al.'s
participants were only given 2$\times$10 minute sessions with the simulation. This leaves very little time for reflection during their decision-making. By contrast, teams completing our study were given as long as they needed, with most teams taking 60--90 minutes to complete a game with 4 decision points. Their findings in a controlled experiment with individuals support what we saw in our dataset of teams collected by the \ac{NPO} in the wild.
Cyber security games are a popular choice for exploring how organization can instill meaningful behavioural change~\cite{denning-2013-control-alt-hack, gondree-2013-security, vykopal-2016-onthedesign, Frey2019TheGoodBadandUgly, bock-2018-kingofthehill, morelock-2018-authenticity,fink2013gamification}.
Other tabletop exercises have been used to help organizations test and refine their response plans (e.g.,\mbox{\cite{bartnes2017challenges,luiijf2010international,high2010promoting}}). Work by Luiijf and Stolk~\mbox{\cite{luiijf2010international}} notes that teams tend to lack long-term focus in decision cycles, much as we saw how teams demonstrate variable levels of reasoning.
Whilst there have been several papers using games and exercises to develop our understanding of how people make cyber security decisions; previous work has studied relatively small numbers of people and under laboratory conditions. Our work, by contrast, is derived from a large dataset of professionals captured \emph{in the wild}.

\section{Discussion}
\label{sec:discussion}

\noindent
{\bf RQ1: What effect does a player's background have on their game performance?}
Our findings in section~\mbox{\ref{sec:findings1:impact of participant background}} suggest that no particular player background, nor organizational sector performed any better than any other within the context of \mbox{\ac{D-D}}. Jalali et al{.}~\mbox{\cite{jalali2019decision}} find a similar trend in their research---indicating that both experts and non-experts employ similar decision-making about cyber security issues. Future work should explore the impact of participant background on cyber security decision-making at a greater level of detail. Our findings assume that the categorization derived with the \mbox{\ac{NPO}} provides an accurate reflection of participants' experience. However, further granularity would be useful to help explore the impact of other performance factors e.g., professional qualifications, extent to which players have responsibility for decision-making in their day-to-day work, and, experience of handling cyber security specific decisions.

{\bf RQ2: What is the effect of diversity of expertise on a team's performance?}
The findings in section~\mbox{\ref{sec:findings1:impact of team make up on performance}} suggest that increasing diversity of expertise in a team has no significant impact on performance within the context of \mbox{\ac{D-D}}.

One results highlight that cyber security decision makers hold biases about how systems \emph{should} be defended, and that these biases are hard to challenge.
This suggests we will continue to make mistakes when defending unfamiliar systems, as how you defend one system may not be suited to another. Future work needs to explore this possibility and consider why these biases have come about.  New technology can help to defend systems, but we also need to look at how this technology can be deployed, and whether it replaces other mitigations.  We need to ensure that the risk decision-makers understand the capabilities \emph{and} limitations of technology, and can apply them appropriately.
To better understand what people do to defend a system, we need to understand their thought processes and the patterns in their reasoning that lead to these choices.

{\bf RQ3: What patterns can be seen in the decision-making processes of players and teams?}
	Section~\mbox{\ref{sec:findings2:why did you buy that?}} highlights a number of broad decision-making patterns that occur during gameplay. For example, teams have a propensity to first purchase the items with which they are most familiar. Despite the experience of the teams studied (as cyber security practitioners), few seemed to use threat modelling in their decision-making; supporting M'Manga et al{.}'s~\mbox{\cite{mmanga2017folk}} theory that risk analysis is rarely a rational process.

Mickens spoke about the \emph{assumptions of technological manifest destiny}, how there is a belief that new kinds of technology ought to \emph{``be deployed as quickly as possible, even if we lack a general idea of how the technology works''.}~\cite{mickens-2018-keynote}.  We saw this in some of the teams studied.  Several teams felt that new technology would solve their problems, they deployed \emph{Firewalls} without discussing what one did or defended because, in their words: \emph{``we always have Firewalls''}. One team (T8) used their training in risk thinking and apply risk methodologies, but went bust while playing as they put investing in \emph{Network Monitoring} above other measures leaving them open to other attacks.
Mickens talked about the need to be skeptical about new technologies and the security they bring---perhaps we need to take this further and not just be skeptical of new technology, but new methodologies, the ability of people to apply them and the ability of experts to know when it is appropriate to apply them?

\section{Conclusion}
\label{sec:conclusion}

Our study explores how player background affects cyber security decision-making, based on a large dataset of 208 teams and 948 different players playing a cyber security tabletop game, collected by a major \ac{NPO}.
We find that, despite a wide range of different backgrounds (including security experts and non-experts), all players tend to play \ac{D-D} similarly, with no particular background playing a statistically better game than any other.
Examining the discussions of 8 teams of cyber security specialists, as they played, reveals that players have biases towards how they approach risk decision making, rarely use threat modeling and favor defending systems with the latest technologies, without understanding exactly how the technology works. Teams rarely apply systematic approaches to decision-making (such as threat modeling) when faced with complex cyber security decisions---something that organizations must address.

\balance
\bibliographystyle{IEEEtran}
\bibliography{references}
\vspace*{-2\baselineskip}
\begin{IEEEbiographynophoto}{Ben Shreeve}
is a Research Associate at University of Bristol.
\end{IEEEbiographynophoto}
\begin{IEEEbiographynophoto}{Joseph Hallett}
is a Senior Research Associate at University of Bristol.
\end{IEEEbiographynophoto}
\begin{IEEEbiographynophoto}{Matthew Edwards}
is a Lecturer at University of Bristol.
\end{IEEEbiographynophoto}
\begin{IEEEbiographynophoto}{Kopo M. Ramokapane}
is a Research Associate at University of Bristol.
\end{IEEEbiographynophoto}
\begin{IEEEbiographynophoto}{Richard Atkins}
is a Cyber Security Advisor at City of London Police.
\end{IEEEbiographynophoto}
\begin{IEEEbiographynophoto}{Awais Rashid}
is a Professor of Cybersecurity at University of Bristol.
\end{IEEEbiographynophoto}

\newpage
\nobalance
\onecolumn

\appendix


The following Tables~11 and 12 report the
uncorrected p-values for pairwise analyses that complement the one-sample
location tests reported in the main body of the paper. No groups were
significantly different at $\alpha = 0.05$ when corrected for the (66) multiple
comparisons. The value $< 0.01$ for \emph{Other+Police} in Table~12 comes closest,
correcting to a $p$ of $0.11$.

\newcommand{\rot}[1]{\multicolumn{1}{c}{\rotatebox{90}{#1}}}
\vspace{2\baselineskip}
\noindent TABLE 11: Uncorrected $p$ values for pairwise t-tests between final score of non-bust teams by player background subgroups.

\begin{center}
\begin{tabular}{lrrrrrrrrrrr}
  \toprule
 & \rot{Board Member} & \rot{Finance} & \rot{Government} & \rot{Legal} & \rot{Manager} & \rot{Other} & \rot{Physical Security} & \rot{Police} & \rot{Risk Analyst} & \rot{Cyber Security} & \rot{Student} \\
  \midrule
  Finance 		& 0.97 &  &  &  &  &  &  &  &  &  &  \\
  Government  		& 0.08 & 0.28 &  &  &  &  &  &  &  &  &  \\
  Legal 		& 0.97 & 0.95 & 0.20 &  &  &  &  &  &  &  &  \\
  Manager 		& 0.71 & 0.81 & 0.16 & 0.82 &  &  &  &  &  &  &  \\
  Other 		& 0.84 & 0.95 & 0.05 & 0.86 & 0.57 &  &  &  &  &  &  \\
  Physical Security 	& 0.79 & 0.81 & 0.45 & 0.82 & 0.95 & 0.71 &  &  &  &  &  \\
  Police 		& 0.95 & 0.94 & 0.08 & 0.99 & 0.75 & 0.78 & 0.81 &  &  &  &  \\
  Risk Analyst 		& 0.72 & 0.82 & 0.10 & 0.84 & 0.95 & 0.55 & 0.92 & 0.76 &  &  &  \\
  Cyber Security 	& 0.56 & 0.81 & 0.02 & 0.68 & 0.33 & 0.72 & 0.58 & 0.49 & 0.25 &  &  \\
  Student 		& 0.72 & 0.82 & 0.13 & 0.74 & 0.55 & 0.81 & 0.62 & 0.68 & 0.56 & 0.95 &  \\
  Technical 		& 0.76 & 0.84 & 0.13 & 0.86 & 0.94 & 0.61 & 0.91 & 0.81 & 0.98 & 0.35 & 0.58 \\
  \bottomrule
\end{tabular}
\end{center}
\vspace{2\baselineskip}

\noindent TABLE 12: Uncorrected $p$ values for pairwise test of proportions between bust status by player background subgroups.

\begin{center}
\begin{tabular}{lrrrrrrrrrrr}
  \toprule
 & \rot{Board Member} & \rot{Finance} & \rot{Government} & \rot{Legal} & \rot{Manager} & \rot{Other} & \rot{Physical Security} & \rot{Police} & \rot{Risk Analyst} & \rot{Cyber Security} & \rot{Student} \\
  \midrule
  Finance 		& 0.42 &  &  &  &  &  &  &  &  &  &  \\
  Government 		& 0.03 & 0.30 &  &  &  &  &  &  &  &  &  \\
  Legal 		& 0.99 & 0.76 & 0.11 &  &  &  &  &  &  &  &  \\
  Manager 		& 0.39 & 0.94 & 0.11 & 0.87 &  &  &  &  &  &  &  \\
  Other 		& 0.63 & 0.20 & 0.01 & 0.65 & 0.11 &  &  &  &  &  &  \\
  Physical Security 	& 0.73 & 1.00 & 1.00 & 0.91 & 1.00 & 0.56 &  &  &  &  &  \\
  Police 		& 0.03 & 0.91 & 0.28 & 0.35 & 0.34 & 0.00 & 1.00 &  &  &  &  \\
  Risk Analyst 		& 0.34 & 0.87 & 0.09 & 0.87 & 1.00 & 0.07 & 1.00 & 0.20 &  &  &  \\
  Cyber Security	& 0.09 & 1.00 & 0.13 & 0.60 & 0.75 & 0.01 & 1.00 & 0.41 & 0.57 &  &  \\
  Student 		& 0.28 & 0.97 & 0.11 & 0.80 & 1.00 & 0.05 & 1.00 & 0.32 & 0.96 & 0.77 &  \\
  Technical 		& 0.74 & 0.25 & 0.01 & 0.71 & 0.20 & 1.00 & 0.59 & 0.01 & 0.15 & 0.04 & 0.13 \\
 \bottomrule
\end{tabular}
\end{center}

\newpage

\noindent TABLE 13: Codebook

\begin{center}
\begin{tabular}{l p{\dimexpr 0.6\linewidth - 2\tabcolsep}}
 \toprule
 Code & Description \\
 \midrule
 No reasoning demonstrated & Teams make a decision without demonstrating any reasoning at all. A suggestion is raised and met with agreement, or with a counter-suggestion.\\
 Minimal reasoning demonstrated & Teams demonstrate minimal reasoning---suggestions may be met with some questioning, typically by only a single participant, but the topic is not pursued for very long.\\
 Extensive reasoning demonstrated & Teams spend a lot of time exploring potential suggestions, multiple lines of questioning are explored by multiple participants. \\
 \addlinespace
 Expertise demonstrated & Participant(s) demonstrate extensive knowledge of a particular aspect of conversation indicative of expertise. \\
 \addlinespace
 Potential impact from choice & Team consider the potential impact from making a particular choice.\\
 Consideration of threats actors? & Team explore the potential threat actors interested in attacking an aspect of the business.\\
 Use of threat modelling? & Teams demonstrate a structured approach for considering threats. Does not have to be a formally recognised model. \\
 Identification/speculation of vulnerabilities & Team explore the potential vulnerabilities that may affect a particular asset or area of the the business.\\
 \addlinespace
 Futuristic tech prioritised & Team chooses to prioritise network monitoring ahead of other alternatives.\\
 Futuristic tech understood? & Team explores network monitoring, but does not appear to understand it entirely. \\
 \addlinespace
 Narrow range of options considered? & Team discusses only a few potential alternatives when make choices. \\
 Wide range of options considered? & Team explores a wide range of potential alternative actions that could be taken. \\
 \addlinespace
 Identifying potential new information & Team sets out to derive information or theories. \\
 Evaluating value of information & Team evaluates validity and value of information. \\
 \addlinespace
 Anticipating future actions & Team takes an action to enable a future action to take place. \\
 \addlinespace
 Cost of investments & Teams evaluate investments based on their cost, often using to prioritise order of investments. \\
 \bottomrule
\end{tabular}
\end{center}

\end{document}